**Title:** Multiparametric quantification and visualization of liver fat using ultrasound

**Authors:** Jihye Baek, PhD[1], Ahmed El Kaffas, PhD[1], Aya Kamaya, MD[1], Kenneth Hoyt, PhD[2,3], Kevin J. Parker, PhD[4]

**Affiliations:** [1]Department of Radiology, Stanford University

[2]Department of Biomedical Engineering, Texas A&M University

[3]Department of Small Animal Clinical Sciences, Texas A&M University

[4]Department of Electrical and Computer Engineering, University of Rochester

**Corresponding author:** Kevin J. Parker, PhD
University of Rochester
724 Computer Studies Building
Box 270231
Rochester, NY 14627, USA
Telephone: +1 (585) 275-3294
Email: kevin.parker@rochester.edu

## Abstract

***Objectives—*** Several ultrasound measures have shown promise for assessment of steatosis compared to traditional B-scan, however clinicians may be required to integrate information across the parameters. Here, we propose an integrated multiparametric approach, enabling simple clinical assessment of key information from combined ultrasound parameters.

***Methods—*** We have measured 13 parameters related to ultrasound and shear wave elastography. These were measured in 30 human subjects under a study of liver fat. The 13 individual measures are assessed for their predictive value using independent magnetic resonance imaging-derived proton density fat fraction (MRI-PDFF) measurements as a reference standard. In addition, a comprehensive and fine-grain analysis is made of all possible combinations of sub-sets of these parameters to determine if any subset can be efficiently combined to predict fat fraction.

***Results—***We found that as few as four key parameters related to ultrasound propagation are sufficient to generate a linear multiparametric parameter with a correlation against MRI-PDFF values of greater than 0.93. This optimal combination was found to have a classification area under the curve (AUC) approaching 1.0 when applying a threshold for separating steatosis grade zero from higher classes. Furthermore, a strategy is developed for applying local estimates of fat content as a color overlay to produce a visual impression of the extent and distribution of fat within the liver.

***Conclusion—***In principle, this approach can be applied to most clinical ultrasound systems to provide the clinician and patient with a rapid and inexpensive estimate of liver fat content.

## 1. Introduction

Due to the increasing prevalence across the globe of fatty liver disease and its many stages along the progression from early steatosis to nonalcoholic steatohepatitis (NASH) and nonalcoholic fatty liver disease (NAFLD), the goal of rapid noninvasive assessment of liver fat has received widespread attention. Ultrasound (US) methods are particularly attractive since they have the potential for rapid, inexpensive implementations even in remote and underserved communities. Active research in US quantification of liver fat has taken at least four different approaches: first, the measurement of a single parameter that trends with increasing fat, for example US attenuation [1, 2]. The disadvantage of this approach includes the presence of cofactors, other than fat accumulation, which can strongly influence any single parameter measurement and create uncertainty [3]. Second, one can derive analytical models based on the biophysics of wave propagation in the liver and solve for the unknown fat content based on accurate measures of fundamental properties such as phase velocity and attenuation [4-6]. Third, one can simply train a machine learning algorithm on a carefully curated set of images, and utilize artificial intelligence (AI) concepts without necessarily incorporating any biophysics [7]. Finally, the approach we take here is a multiparametric analysis leading to a distillation of the most important set of measures and their most simplified combination in order to accurately quantify liver fat. The combination of independent parameters allows for an accounting of the effects of cofactors, and a tighter correlation with the key independent measure of fat, in this case the widely used magnetic resonance imaging-derived proton density fat fraction (MRI-PDFF) [8]. Our multiparametric approach can also be deployed with machine learning techniques such as the support vector machine (SVM) [9, 10] where a training set using our measures can produce an accurate segmentation of diseases and stages of diseases in a multidimensional, multiparametric space [11-13]

Accordingly, this paper is organized in stages to consider a broad set of parameters that can be measured in the liver using a clinical scanner (Philips EPIQ 7, Philips Healthcare, Bothell, WA, USA) on 30 patients who are being evaluated for steatosis. This includes US- and elastography-related measures. Next, the individual parameters and combinations of parameters are assessed for their correlation with the independent measure of liver fat using MRI-PDFF. Several of these parameters and correlations have been reported previously from the original study conducted at Stanford University [14], and in this follow-up we add additional measures related to the H-scan analysis and the power law framework leading to Burr parameters for speckle characterization [15], Then, an exhaustive accounting of all possible combinations of parameters is conducted to identify if any subset of these can produce a highly effective predictor of liver fat. Principal component analysis (PCA) is used to further simplify the combinations of parameters. A limited subset of as few as four parameters (in principle, obtainable by most modern US scanners) is found to produce strong correlation against fat fraction assessed by MRI-PDFF and area under the curve (AUC) approaching 1.0 for classification by steatotic score of zero vs. all higher scores. Finally, in addition to the predictive and classification uses of this approach, using the strongest few parameters we are able to make local predictions of fat content and display this information as color overlay images, producing an immediate visual impression of the amount and location of the fat within each liver. These quantitative imaging results are referred to as US fat fraction (USFF) images.

Taken together, these analyses demonstrate an effective and efficient means to generate liver fat estimates that are strongly correlated with MRI-PDFF assessments, and tightly linked to steatosis scores, and can also produce images that convey an immediate impression of the quantity and distribution of fat within the liver.

## 2. Methods

### 2.1 Study design

We studied *in vivo* human subjects with healthy livers and suspicious or confirmed NAFLD having at least 1 associated risk factor of obesity, diabetes, or hypertension. The Stanford University School of Medicine Review Board approved this prospective Health Insurance Portability and Accountability Act–compliant study. The study screened 211 patients who underwent magnetic resonance imaging (MRI) liver scanning at Stanford University Medical Center but excluded patients for specific reasons; more detailed patient information can be found in the previous study with the same patient dataset [14]. Overall, just 30 patients were eligible with available US radiofrequency (RF) data, quantitative US (QUS) parameters, US shear wave measures, and reference MRI-PDFF measures acquired from the same MRI system. The patients had reference steatosis MRI-PDFF ranging from 1.25 % to 42.8 % (14.1 % ± 11.3 %).

The previous study [14] provided 8 measures incorporated into this study, related to ultrasound measures of: hepatorenal index (HRI), Nakagami analysis, spectral slope, spectral intercept, midband fit, plus elasticity measures of shear wave speed (SWS), shear wave viscosity (SWV), and shear wave dispersion (SWD at 100-150 Hz, 150-200 Hz, and 100-200 Hz), and this study extracted 5 additional measures (H-scan, attenuation, Burr lambda, Burr b, B-scan intensity). A total of 13 measures were utilized for multiparametric analysis, which evaluated all possible combinations and ranked them based on correlation coefficient and the AUC with reference to MRI-PDFF measures. This evaluation determined the best subset to quantify liver steatosis, producing a combined parameter which was utilized to visualize steatosis color overlay on traditional B-mode images.

## 2.2 Ultrasound parameters

### *2.2.1 Ultrasound acquisition*

The patients were US-scanned using the Philips EPIQ 7 system (Philips Healthcare, Bothell, WA) equipped with a C5-1 convex transducer. Up to 12 views of US images were acquired for liver and kidney in each patient. The liver images were used to measure all US parameters, whereas the kidney scans were only used to measure HRI. These scans saved US RF data for our analysis. Following the B-scan, US shear wave elastography (SWE) was performed, and SWE parameters were measured. Example views of B-mode kidney and liver and SWE are illustrated in **Figure 1**.

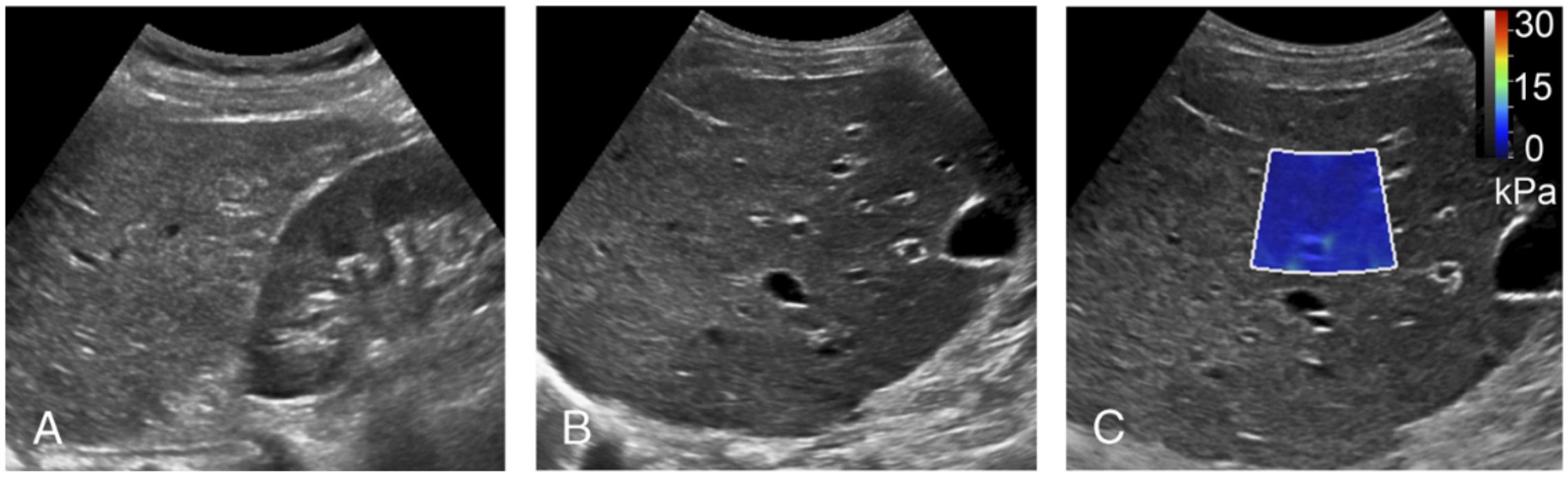

**Figure 1** Ultrasound (US) acquisitions with different views and techniques. (a) B-mode hepatorenal view for quantifying the hepatorenal index (HRI), (b) B-mode liver view for the other B-mode parameters extracted using radiofrequency (RF) data, (c) Shear wave elastography (SWE) mode for shear wave propagation parameters. Permission to re-use in process [14].

### *2.2.2 H-scan*

Analyzing the RF data enables H-scan and attenuation estimation based on frequency information. The H-scan is a matched filter analysis, which is capable of characterizing tissue properties. As a preprocessing for H-scan, attenuation correction was applied to the RF data [16] since the US attenuation effect causes frequency down-shift along with depth, resulting in a red-shift of H-scan. We multiplied $e^{\alpha f_{tx} x_z}$ to Fourier-transformed RF data in each zone ($z$) where the overall depth was divided into 10 zones, with each zone having a length of 1.6 cm, and $x_z$ is the center depth of zone $z$. $\alpha$ and $f_{tx}$ are the attenuation coefficient ($\alpha$ = 0.5 dB/MHz/cm) and transmission frequency 3 MHz, respectively. The matched filtering was applied to the attenuation-compensated RF data which then produced convolved signals highlighting specific scatterer sizes related to peak frequencies of the filters. In this study, we utilized 256 Gaussian filters with peak frequencies between 1.5 MHz and 4.5 MHz, with equivalent difference of 23.6 kHz. Each pixel had 256 convolution values, and a maximum of the convolution can be selected. The maximum convolution had a corresponding Gaussian filter index between 1 and 256. The indices of maximum convolved signals were color-coded to produce H-scan images using the H-scan colormap in **Figure 2**.

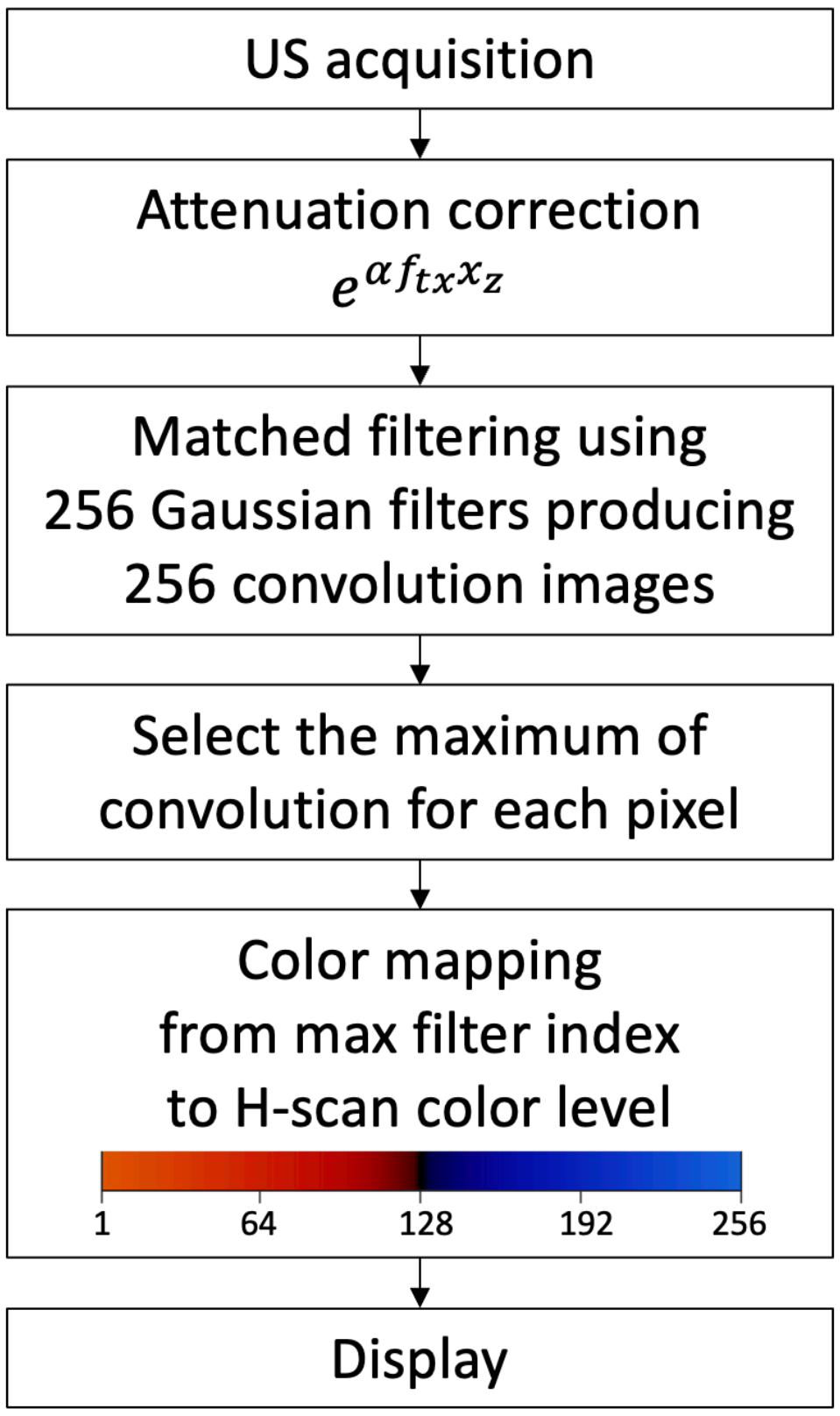

**Figure 2** H-scan flow chart.

The lower peak frequency indices mapped into lower color levels showing more red color represent relatively larger US scatterers, whereas the higher frequencies/color levels showing more blue represent smaller US scatterers. The H-scan color levels were used as an US parameter characterizing US scatterer sizes, which can vary due to pathological changes.

*2.2.3 Attenuation estimation*

The attenuation coefficient can be estimated using the H-scan procedure without applying attenuation correction (RF data preprocessing for H-scan). Thus, raw RF data without attenuation correction were used as the input. The raw RF data were matched filtered, which can provide peak frequency components along with depth ($\hat{f}_p(x)$). Then, estimated attenuation coefficient ($\hat{\alpha}(x)$) can be calculated:

$$\hat{\alpha}(x) = -\frac{\hat{f}_p(x) - f_{tx}}{x\sigma^2}, \tag{1}$$

where $\sigma$ is bandwidth of the frequency spectrum; more details of this estimation method can be found in Baek *et al.* [10]. $\hat{\alpha}(x)$ was averaged along with depth, and the average was used as one of our US parameters.

*2.2.4 Burr and Nakagami parameters*

Derived from speckle theory tracing back to Rayleigh's 1880 derivations [17], a number of treatments of speckle conclude with a probability density function (PDF) including an exponential or Gaussian tail. The Nakagami parameter ($m$) is one of these, comprising a two-parameter distribution given as:

$$P(A) = \frac{2m^m A^{2m-1}}{\Gamma(m)\Omega^m} \exp\left(-\frac{m}{\Omega} A^2\right) U(A), \tag{2}$$

where $\Gamma$ and $U$ are the gamma and unit step functions, respectively, $A$ is the echo amplitude, $\Omega$ is the scaling parameter defined as $\Omega = E(R_{env}^2)$, where $E$ denotes the statistical mean and $R_{\text{env}}$ is the echo envelope signals [18].

A more recent alternative is derived from the premise that there exists in tissue a multiscale, power law distribution of scattering sites, leading to a Burr distribution of speckle amplitudes. This can be characterized as a "long tail" distribution, and is given as a two-parameter PDF by:

$$P(A) = \frac{2A(b-1)}{\lambda^2\left[\left(\frac{A}{\lambda}\right)^2 + 1\right]^b}, \tag{3}$$

where $\lambda$ and $b$ are the two Burr parameters to be estimated. $\lambda$ is a scale factor related to echo amplitude and gain and $b$ is a power law exponent related to scatterer distribution. These two distributions can have similar shapes on a linear histogram plot, the major difference is present in the tails, exponential vs. power law.

*2.2.5 B-scan intensity and hepatorenal index*

Analyzing B-mode echogenicity can produce B-scan intensity and HRI. To measure B-scan intensity, the saved RF data were processed to IQ-data and then envelope data. The envelope data amplitudes were averaged within a region of interest (ROI) including only soft tissues after excluding vasculature. HRI was calculated by setting two boxes in the liver and kidney as previously described [14]:

$$\text{HRI} = \frac{\text{mean liver echogenicity}}{\text{mean kidney echogenicity}}. \tag{4}$$

*2.2.6 Shear wave parameters*

SWE imaging sequences produced the six shear wave related parameters for this study: elasticity (shear modulus); a fit of data to a Voigt rheological model (Voigt-viscosity) [19] or to an experimental hybrid model (WE-viscosity) [20]; dispersion from the range of 100-150 Hz; dispersion 150-200 Hz; and dispersion 100-200 Hz. These parameters were measured as part of the previous study; the extraction details are available in Pirmoazen *et al*. [14].

## 2.3 MRI-PDFF

As an independent reference measure for the patient data set, MRI-PDFF was used. Standard MRI scanning for the patients was performed at Stanford Medical Center by a radiologist with 10-years of experience using a 3.0 T scanners (Discovery MR750, GE Medical System, Waukesha, WI) within 14 days of the US scanning. 55% of the patients underwent both MRI and US scanning on the same day, and the average number of days between the MRI and US was 3.93 $\pm$ 5.25 days. The MRI-PDFF measures were obtained using a 6-echo 3-dimensional (3D) spoiled gradient recalled echo sequence (IDEAL-IQ). The radiologist created ROIs in segments 5 – 8 of the right hepatic lobes. The four measurements were averaged and used as the reference MRI-PDFF. Utilizing the MRI-PDFF measures, hepatic steatosis was staged using the following MRI-PDFF cutoffs: Normal (S0) < 5% < S1 < 10% < S2 < 20% < S3.

## 2.4 Multiparametric analysis

*2.4.1 Feature selection*

Multiparametric analysis was performed to select the best combination which accurately estimated fat fraction by extracting and combining information from the individual features. We investigated all possible combinations with two categories. The first category included all combinations from the 13 parameters from RF data and SWE, and the second included only parameters extracted from B-mode, without SWE. Since measuring SWE requires SWE transmissions in addition to the B-mode sequence, we evaluated the performance of the simpler protocol only with B-mode. The first category has 8191 possible unique combinations:

$$\binom{13}{13}+\binom{13}{12}+\binom{13}{11}+\cdots+\binom{13}{2}+\binom{13}{1}=\sum_{k=1}^{13}\binom{13}{k}=8191, \tag{5}$$

where $\binom{13}{k}$ denotes binomial coefficient, representing the possible number of combinations when selecting $k$ parameters from among 13 parameters. In the same way, the second category had 127 possible combinations:

$$\sum_{k=1}^{7}\binom{7}{k}=127, \tag{6}$$

where $\binom{7}{k}$ represents possible number of combinations when selecting $k$ parameters among seven.

The performance of all possible combinations were evaluated using linear ($R$) and Spearman's ($R_s$) correlation coefficients and the AUC with three different thresholds: (1) S0 vs. S1/S2/S3, (2) S0/S1 vs. S2/S3, (3) S0/S1/S2 vs. S3. To assess the performance considering the total of five evaluations, including correlation coefficients and AUCs, we propose a combined metric (CM):

$$\mathrm{CM}=0.5\cdot\left(\frac{\mathrm{R}+\mathrm{R_s}}{2}\right)+0.5\cdot\left(\frac{\mathrm{AUC_{S0\ vs.\ S1S2S3}}+\mathrm{AUC_{S0S1\ vs.\ S2S3}}+\mathrm{AUC_{S0S1S2\ vs.\ S3}}}{3}\right), \tag{7}$$

where $\mathrm{AUC_{S0\ vs.\ S1S2S3}}$ is the AUC with the threshed between S0 and S1, $\mathrm{AUC}_{S0S1\ \mathrm{vs.}\ S2S3}$ is the AUC with the threshed between S1 and S2, and $\mathrm{AUC_{S0S1S2\ vs.\ S3}}$ is the AUC with the threshed between S2 and S3. The metric has 50% weight from correlation coefficients and 50% weight from AUCs. When $R$ and $R_s$ were used to evaluate individual parameters' performance, p-values less than 0.001 were considered statistically significant.

The metric (eqn (7)) was calculated for all 8191 and 127 feature combinations for the first (all US parameters) and second (B-mode parameters excluding SWE) categories of parameters, respectively, and we found the best performing combinations from each category and compared the two.

*2.4.2 Multiparametric quantification and imaging*

Once we determined the best performing parameter combinations, multiparametric analysis combined information from the selected parameters using principal component analysis (PCA). The first principal component (PC1) can be considered to be a combined, or synthesized, single parameter, whose performance was evaluated using the correlation coefficients and AUCs and compared with the 13 individual parameters. The combined parameter was utilized to estimate USFF, and the estimated USFF was color-coded using the USFF color bar (**Figure 3**) and overlaid onto B-mode images.

**Figure 3** depicts the USFF imaging method. MRI-PDFF is used as a reference standard for fat quantification. US parameters are measured, and feature selection (Section 2.4.1.) finds the best performing feature combination. PCA combines the parameters and provides a combined parameter of PC1. The strong correlation between PC1 and MRI-PDFF produces a linear fit line. This overall approach to quantifying hepatic steatosis utilizes information from feature selection and PCA. Thus, for prediction, we only measure the selected features, excluding features that do not contribute to better performance of the multiparametric analysis for steatosis evaluation. The parameters are sent to PCA using the PCA matrix calculated from training. Then, the combined parameter, PC1, is calculated from the measurements and the PCA matrix. Using the linear fit line from training, we can estimate USFF from the PC1. The USFF is color-codded and overlaid onto B-mode images.

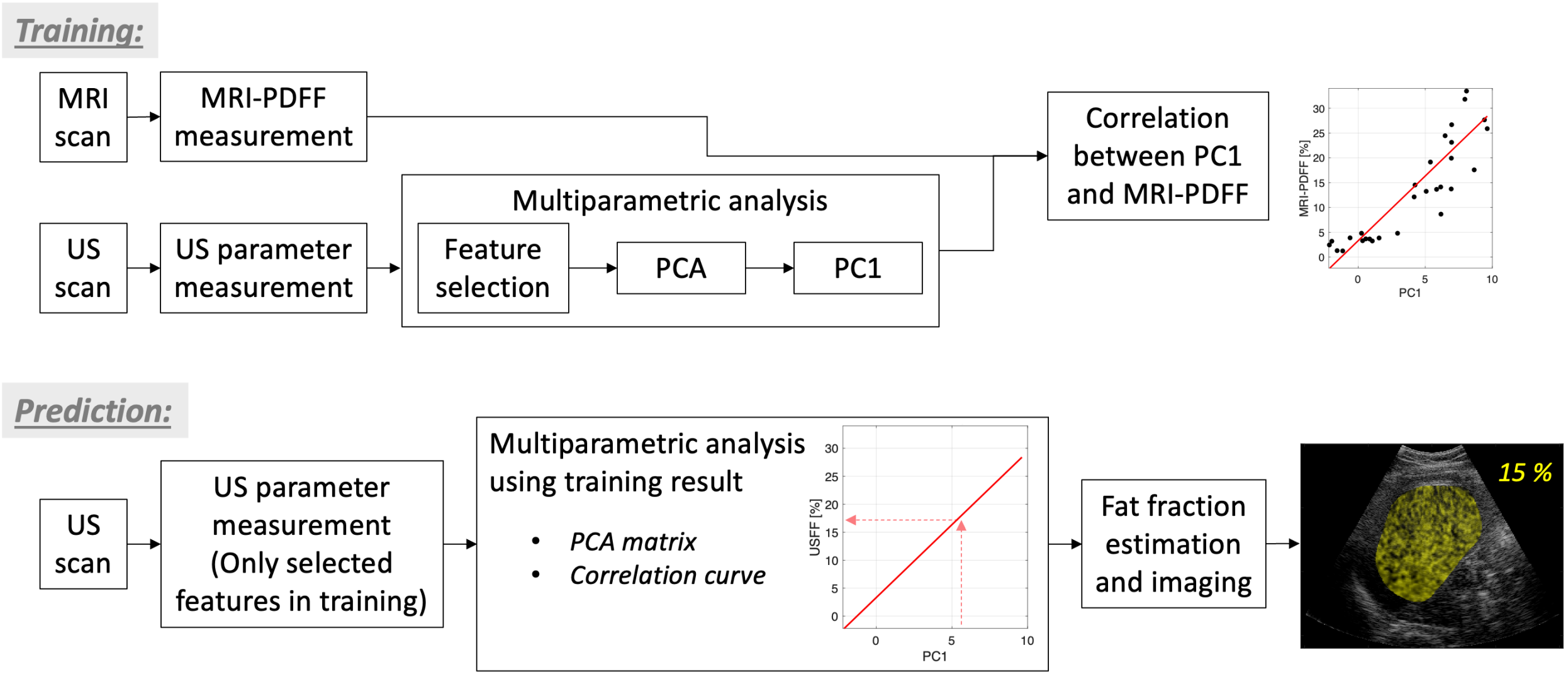

**Figure 3** Multiparametric imaging flow chart. MRI-PDFF = magnetic resonance imaging-derived proton density fat fraction, PCA = principal component analysis, PC1 = the first principal component.

## 3. Results

### 3.1 Individual ultrasound parameters

The 13 US individual parameters were measured and compared using MRI-PDFF (**Figure 4**). Both H-scan and attenuation measures showed the highest correlation of $|R_s|= 0.90$ ($p < 0.001$) with MRI-PDFF. The Nakagami parameter also showed high correlation, with $|R_s| = 0.88$ ($p < 0.001$). Further, the Burr lambda, Nakagami, and HRI correlated well with MRI-PDFF with correlation higher than $R_s = 0.70$ ($p < 0.001$). However, all SWE parameters failed to show significant correlation with MRI-PDFF; for all SWE parameters, $|R_s| < 0.50$ ($p > 0.01$).

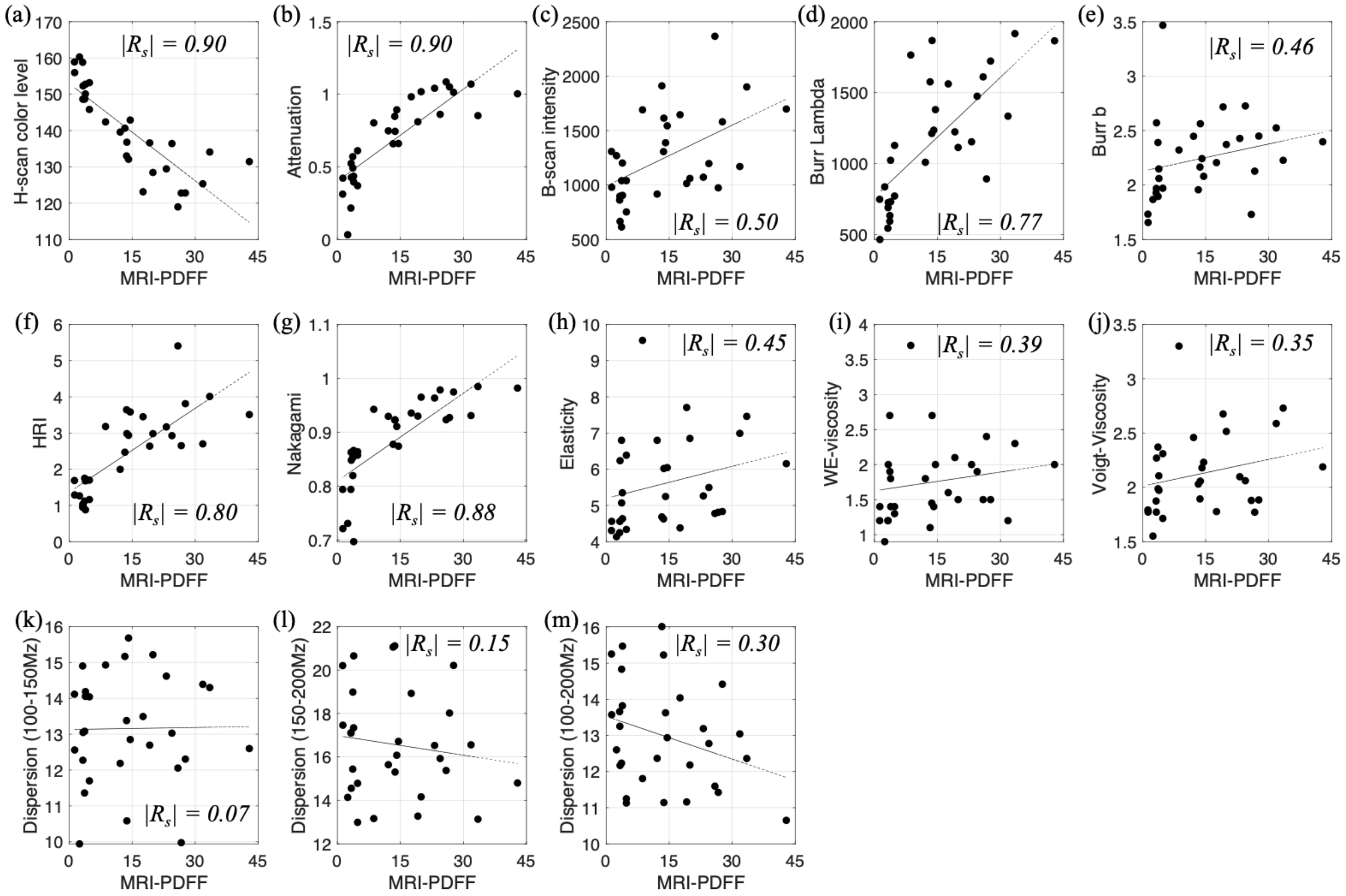

**Figure 4** Fat quantification performance of individual parameters was compared with reference MRI-PDFF. (a) H-scan color level, (b) attenuation coefficient, (c) B-scan envelope intensity, (d) Burr lambda, (e) Burr b, (f) hepatorenal index (HRI), (g) Nakagami, (h) SW elasticity [kPa], (i) WE viscosity, (j) Voigt-viscosity, (k) SW dispersion 100-150 Hz, (l) dispersion 150-200 Hz, and (m) dispersion 100-200 Hz. SW = shear wave.

Of all the 13 individual parameters, H-scan and attenuation outperformed the others. These two parameters were extracted from our frequency-domain signal processing and thus frequency analysis outperformed echogenicity- and shear wave-based analysis to assess hepatic steatosis.

Our combined metric (CM, eqn (7)), obtained from correlation coefficients and AUCs, evaluated the performance of individual parameters (**Figure 5**). H-scan and attenuation performed the best with CM greater than 0.9, as provided in **Figure 5**. Also, Nakagami, HRI, and Burr lambda showed high performance (CM > 0.8). However, shear wave parameters provided relatively poor steatosis assessment.

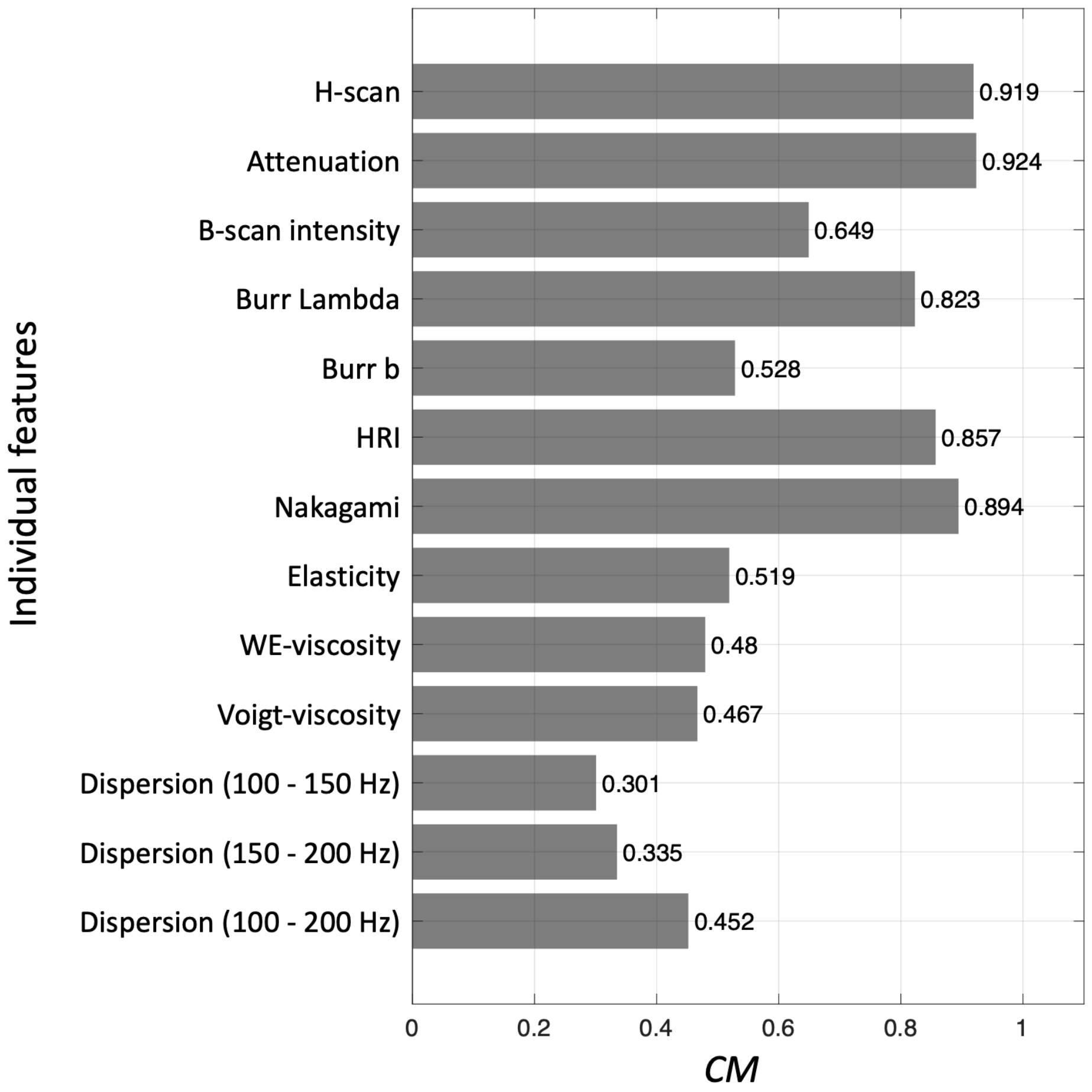

**Figure 5** Performance of individual parameters evaluated by our combined metric (CM, eqn (7)) utilizing correlation coefficients and AUCs. HRI = hepatorenal index.

### 3.2 Multiparametric quantification

**Figure 6** shows performance for all possible parameter combinations when including all 13 parameters (**Figure 6(a)**) and then including only B-mode derived parameters without SWE (**Figure 6(b)**). **Figures 6(c)** and **6(d)** show the highest 30 combinations for the two categories. The best performing combination for both categories is (H-scan, attenuation, Burr lambda, Nakagami), with CM = 0.938. This combination does not include any SWE parameter and, thus, SWE is not essential for the overall assessment. The parameters extracted from B-mode alone can accurately assess hepatic steatosis, resulting in a strong agreement with MRI-PDFF.

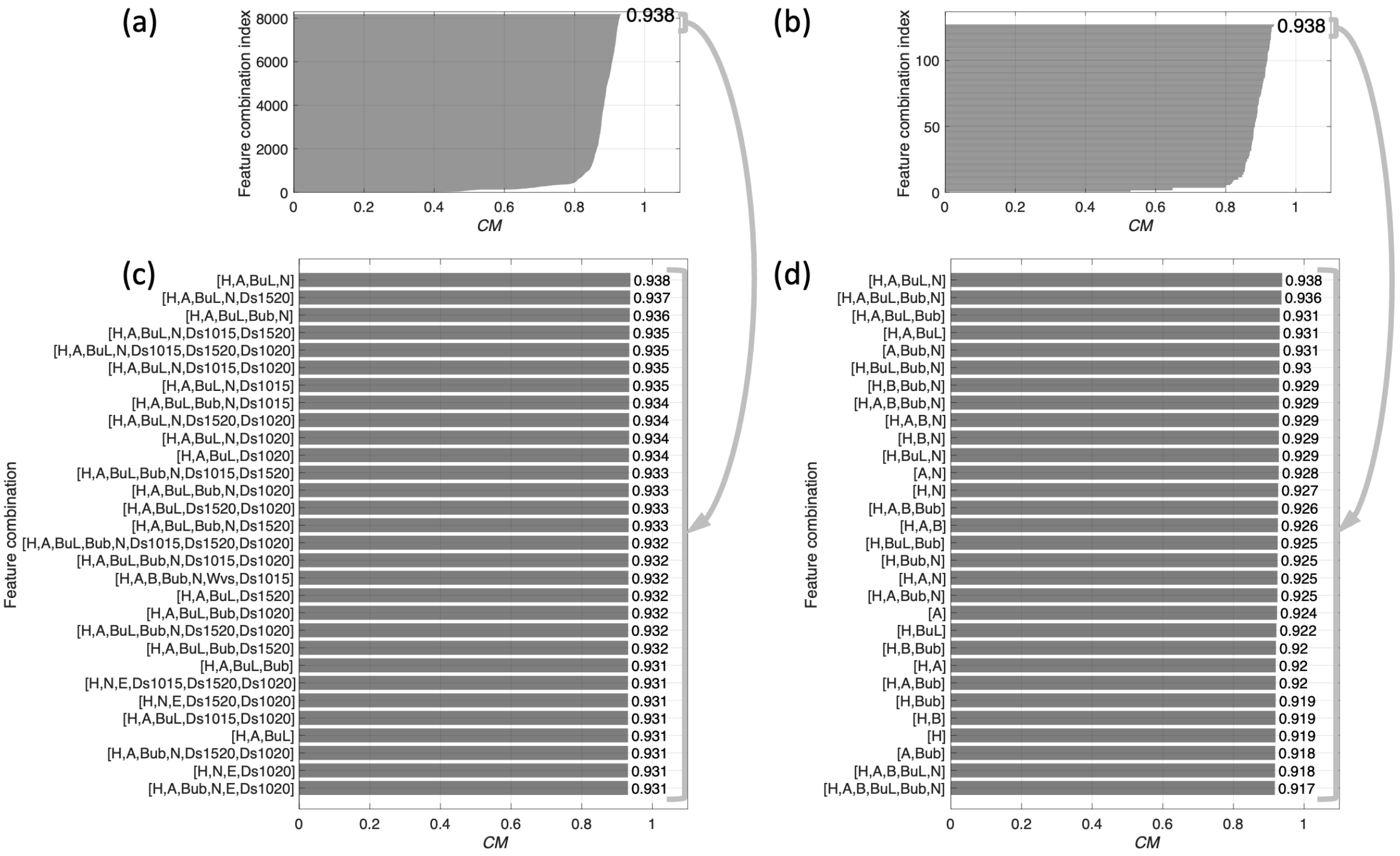

**Figure 6** Feature selection results when (a) including all US parameters and (b) including only B-mode parameters but excluding SWE parameters. (c) and (d) include the best 30 combination for (a) and (b), respectively. H = H-scan, A = attenuation coefficient, B = B-scan intensity, BuL = Burr lambda, Bub = Burr b, N = Nakagami, E = SW elasticity, Wvs = WE-viscosity, Ds1015 = SW dispersion 100-150 Hz, Ds1020 = SW dispersion 100-200 Hz, Ds1020 = SW dispersion 150-200 Hz.

**Figure 7** provides more information on the best performing feature combination's performance. **Figure 7(a)** shows the correlation plot between MRI-PDFF and our CM, showing strong correlation ($R_s = 0.93$). **Figure 7(b)** provides receiver operating characteristic (ROC) curves with 3 different thresholds: (1) S0 vs. S1, S2, S3 (AUC = 1.00); (2) S0, S1 vs. S2, S3 (AUC = 0.98); (3) S0, S1, S2 vs. S3 (AUC = 0.96). **Figure 7(c)** provides contribution from each parameter when we calculated our CM. It indicates that H-scan contributed more than the other parameters. The contribution from frequency-related parameters (H-scan and attenuation coefficient) is higher than echogenicity-based analysis (Burr lambda and Nakagami). Furthermore, the results of these combined parameters (**Figures 6-7**) outperformed any individual parameters (**Figure 5**), indicating that multiparametric analysis resulted in improved steatosis assessment.

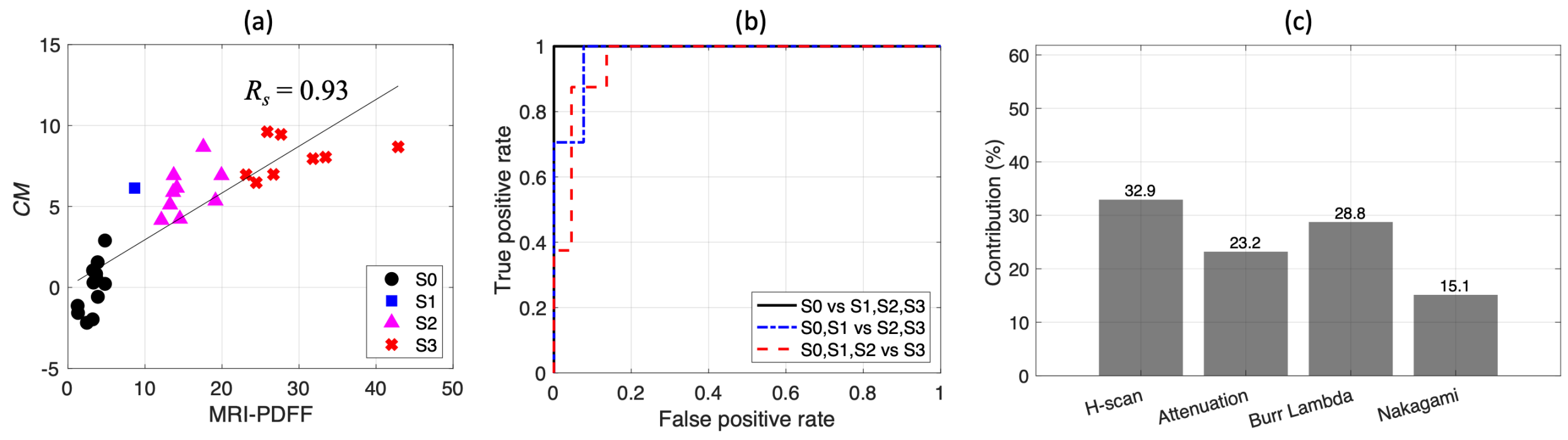

**Figure 7** Best performing feature combinations found using multiparametric analysis. (a) shows correlation coefficient ($R_s$ = 0.93) between our combined metric (CM) and MRI-PDFF. (b) shows ROC curves with different cutoffs. (c) provides contribution from each parameter.

## 3.3 Multiparametric imaging: USFF

Multiparametric analysis produced a combined parameter from selected individual features: H-scan, attenuation, Burr lambda, Nagakami. The combined parameter for each pixel was color-coded and overlaid onto B-mode imaging (**Figure 8**). **Figure 8** compares patient images from B-scan, H-scan, H-scan, and multiparametric analyses, without and with segmentation, with increasing fat accumulated from left to right. The figures show that fat accumulation causes intensity transition from dark to bright for B-scan and color transition from blue to red for H-scan. Multiparametric imaging illustrates more yellow-overlaid tissues and transition from dark to bright yellow as fat increases from left to right: for normal or low-stage steatosis, traditional B-mode images are more visible, but for steatotic cases, there are more highlighted fatty tissues. Although all imaging methods are capable of illustrating fat accumulation in liver, H-scan and multiparametric imaging tend to display the changes better than B-scan. Furthermore, multiparametric imaging can also provide quantitative fat percentages with color and color bar, whereas B-mode and H-scan only provide relative changes.

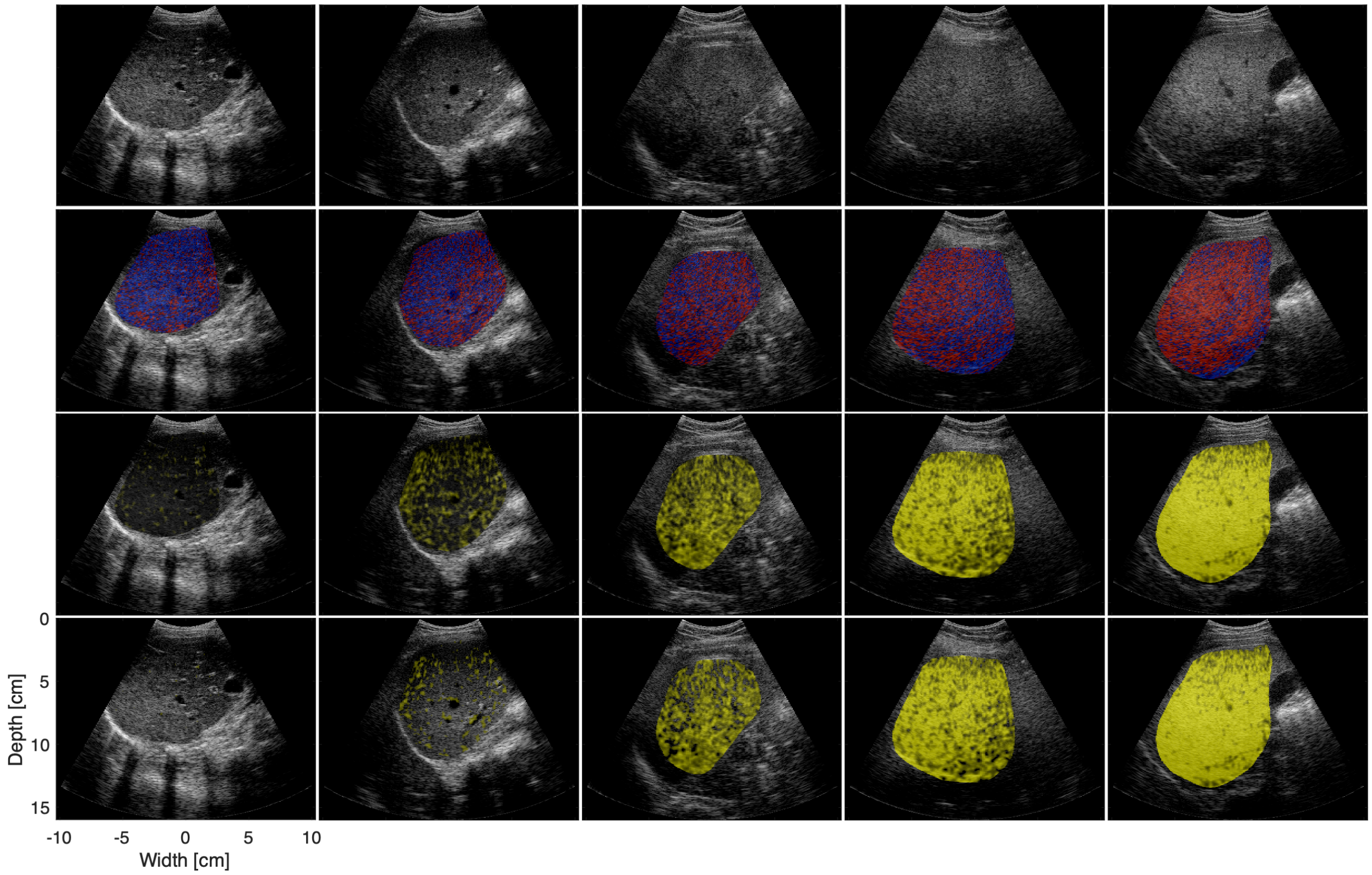

**Figure 8** Illustration of fat accumulation: from top to bottom, B-scan, H-scan, multiparametric imaging, and multiparametric imaging segmentation. MRI-PDFF from left to right: 2.5, 4.8, 12.1, 19.9, and 25.9 %.

## 4. Discussion

This study addresses two levels of questions about diagnostic US. The larger question is: how can we integrate a wide range of measurable parameters to accurately stage disease and report to the clinician? The narrower question is: what subset of measurable parameters is best able to quantify liver steatosis? Our approach to these questions included a comprehensive look at 13 individual measures related to US and shear wave propagation in the liver and then examined a fine-grain, exhaustive study of all possible combinations of subsets of the measured parameters. We then applied a set of techniques to integrate the best measures, classify results, and produce images with color overlays to indicate the extent and severity of the liver fat.

The leading parameters for assessing steatosis included fundamental properties related to power law relations that underlie medical US. Backscatter, attenuation, and the Burr distribution of speckle all contributed to a strong multiparametric assessment of fat, and all are related at a deep level of biophysics to power laws stemming from the multiscale distribution of structures within the soft tissues [21]. Also, the hepatorenal index, a time-honored comparison of the B-scan brightness of the liver compared with the kidney, worked well in our study, presumably implying that in the 30 subjects, kidney echogenicity was relatively unchanged by the progression of fatty liver, NASH, and NAFLD. This may not always be the case in patients with comorbidities and will have to be examined in larger populations. The speckle distribution characterized by the Nakagami distribution is also found to be a valuable contributor. This distribution bears a strong resemblance to the Burr distribution except at the high intensity tail (the Burr distribution is a "long tail" distribution unlike most earlier models tracing back to Rayleigh), yet the tail represents a minimal percent of the data within small ROIs, so it may be a minor distinction in practical applications. We note that the elastography-related measures were not strong contributors to correlations against the independent MRI-PDFF measurement of liver fat. This is likely to be related to the influence of cofactors that can be present and can also influence shear wave propagations [3]. As a practical matter, this relatively poor correlation implies that clinical B-scan platforms, properly arranged for basic US measures, would not require additional elastography capabilities (for example, high intensity push-pulse capabilities) to extract an accurate estimate of liver fat. This simplifies the requirements for implementing our approach.

Our multiparametric approach enabled more accurate assessment of hepatic steatosis for human subjects compared to the individual US parameters. Our *in vivo* animal study [22] concluded that our analysis enhanced steatosis evaluation performance compared to individual US parameters and further achieved better performance than MRI-PDFF. We expect to see our multiparametric analysis outperforming MRI-PDFF for human subjects once we compare US and MRI data with a biopsy reference.

Our approach combined the multiple parameters using linear PCA, but as shown in **Figure 7 (a)**, the scatter plot showed non-linearity. The linear and non-linear correlation coefficients are $R = 0.87$ and $R_s = 0.93$, respectively; the non-linear Spearman's correlation coefficient is higher than the linear coefficient. Thus, non-linear feature combination methods could enable more accurate steatosis estimation. For example, non-linear PCA for mapping from MRI-PDFF to PC1 and mapping from PC1 to fat percentages in **Figure 3** can be considered. Further, our previous study [23] suggested a new non-linear method utilizing a non-linear hyperplane obtained from the support vector machine, which achieved higher diagnostic accuracy compared to linear combinations of PCA and inner-product.

We scanned the livers multiple times for the same patient (mean: 3.5, minimum: 3, maximum: 7). **Figure 9** displays example scans for two patients, one with normal liver and one with steatotic liver. The liver sections are slightly different between scans, but show comparably highlighted portions of yellow between the scans. For the normal and steatotic case, USFF is 5.78% (standard deviation, SD = 1.02) and 27.3 % (SD = 1.44), respectively. For all 30 patients, averaged SD is 1.38 %, which proves our estimator's reliable measurements from scan to scan.

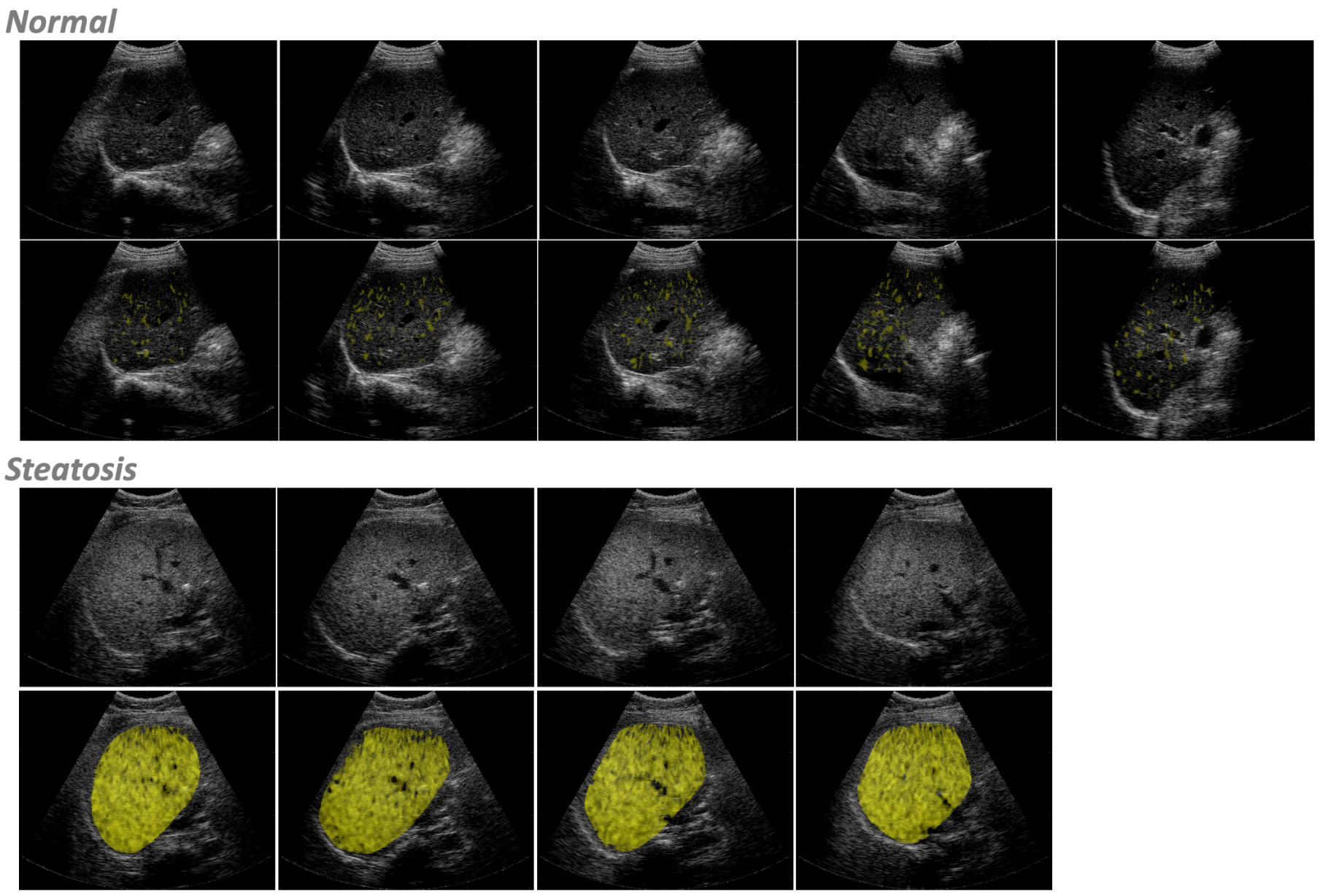

**Figure 9** Multiple scans from normal and steatotic patient.

Limitations of the study include the limited size of the patient population and the difficulty of extracting reliable measures from the most distal portions of the liver. In longer propagation paths, on a scale greater than 10 cm, the issues of attenuation, gain, noise, and beam diffraction all accumulate to create larger uncertainty in the measurement of parameters and the proper correction for depth-dependent effects. Nonetheless, we have reasonable stability in our multiparametric synthesis and USFF images throughout this population, as seen in **Figures 8** and **10**.

We further provide a possible display mode of our multiparametric imaging when it is applied to commercial machines. Overall analysis of this study investigated the liver segmented ROIs for more accurate evaluation for this hepatic steatosis study. However, for routine use, clinicians can set the ROI box as shown in **Figure 10**, simply indicating the angle and depth range to be investigated.

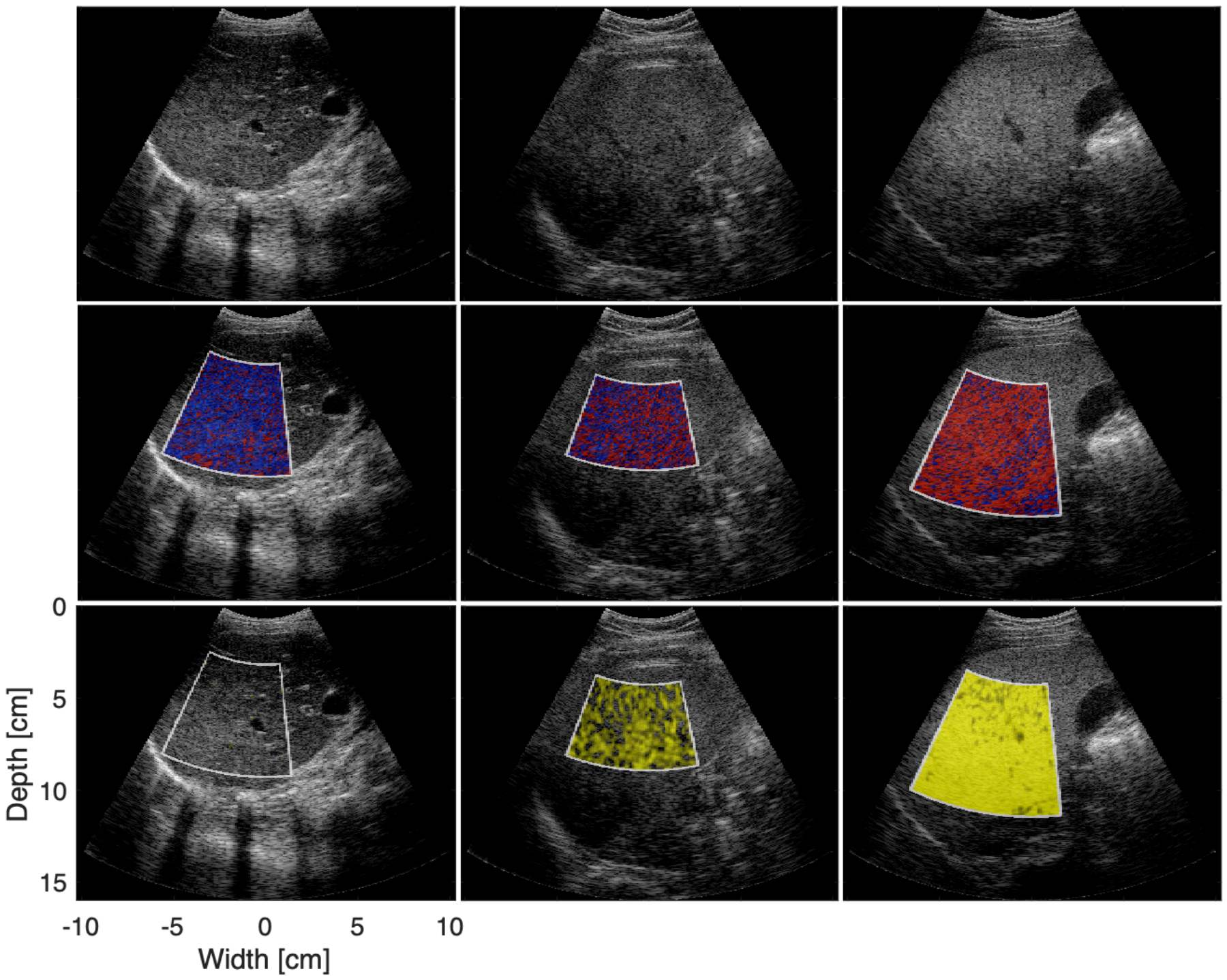

**Figure 10** Possible display mode of our proposed multiparametric imaging for commercial machines, which requires a simple ROI selection for clinicians. Top to bottom: B-scan, H-scan, and multiparametric imaging. Left to right: livers with gradations of normal to severe steatosis.

# 6. Conclusions

In summary, the results here show promise for US-related measures (without the need for elastography measures) that can be integrated into a reliable measure of liver fat and simultaneously can produce color overlay images that provide a visual depiction of the extent of the fat within the liver. These capabilities are germane to the broader goal of providing rapid, inexpensive assessment of steatosis and liver disease to the global population.

# Data statement

Requests for data can be made to the corresponding author.

## Declaration of competing interest

The authors have no competing interests.

## Acknowledgments

This work was supported by National Institutes of Health grant R01DK126833.

## References

1. Ajmera, V. and R. Loomba, *Imaging biomarkers of NAFLD, NASH, and fibrosis.* Mol Metab, (2021). **50**: p. 101167.
2. Petzold, G., *Role of ultrasound methods for the assessment of NAFLD.* J Clin Med, (2022). **11**(15).
3. Poul, S.S. and K.J. Parker, *Fat and fibrosis as confounding cofactors in viscoelastic measurements of the liver.* Phys Med Biol, (2021). **66**(4): p. 045024.
4. Parker, K.J., J. Ormachea, M.G. Drage, H. Kim, and Z. Hah, *The biomechanics of simple steatosis and steatohepatitis.* Phys Med Biol, (2018). **63**(10): p. 105013.
5. Parker, K.J. and J. Ormachea, *The quantification of liver fat from wave speed and attenuation.* Phys Med Biol, (2021). **66**(14): p. 145011.
6. Ormachea, J. and K.J. Parker, *A preliminary study of liver fat quantification using reported ultrasound speed of sound and attenuation parameters.* Ultrasound Med Biol, (2022). **48**(4): p. 675-684.
7. Alshagathrh, F.M. and M.S. Househ, *Artificial intelligence for detecting and quantifying fatty liver in ultrasound images: A systematic review.* Bioengineering (Basel), (2022). **9**(12).
8. Pirmoazen, A.M., A. Khurana, A. El Kaffas, and A. Kamaya, *Quantitative ultrasound approaches for diagnosis and monitoring hepatic steatosis in nonalcoholic fatty liver disease.* Theranostics, (2020). **10**(9): p. 4277-4289.
9. Baek, J., T.A. Swanson, T. Tuthill, and K.J. Parker. *Support vector machine (SVM) based liver classification: fibrosis, steatosis, and inflammation*. in *2020 IEEE International Ultrasonics Symposium (IUS)*. 2020.
10. Baek, J., S.S. Poul, T.A. Swanson, T. Tuthill, and K.J. Parker, *Scattering signatures of normal versus abnormal livers with support vector machine classification.* Ultrasound Med Biol, (2020). **46**(12): p. 3379-3392.

11. Baek, J. and K.J. Parker. *Disease-specific imaging with H-scan trajectories and support vector machine to visualize the progression of liver diseases*. in *2021 IEEE International Ultrasonics Symposium (IUS)*. 2021.
12. Baek, J., L. Basavarajappa, K. Hoyt, and K.J. Parker, *Disease-specific imaging utilizing support vector machine classification of H-scan parameters: assessment of steatosis in a rat model.* IEEE Trans Ultrason Ferroelectr Freq Control, (2022). **69**(2): p. 720-731.
13. Baek, J. and K.J. Parker, *H-scan trajectories indicate the progression of specific diseases.* Med Phys, (2021). **48**(9): p. 5047-5058.
14. Pirmoazen, A.M., A. Khurana, A.M. Loening, T. Liang, V. Shamdasani, H. Xie, A. El Kaffas, and A. Kamaya, *Diagnostic performance of 9 quantitative ultrasound parameters for detection and classification of hepatic steatosis in nonalcoholic fatty liver disease.* Invest Radiol, (2022). **57**(1): p. 23-32.
15. Baek, J., S.S. Poul, L. Basavarajappa, S. Reddy, H. Tai, K. Hoyt, and K.J. Parker, *Clusters of ultrasound scattering parameters for the classification of steatotic and normal Livers.* Ultrasound Med Biol, (2021). **47**(10): p. 3014-3027.
16. Parker, K.J. and J. Baek, *Fine-tuning the H-scan for discriminating changes in tissue scatterers.* Biomed Phys Eng Express, (2020). **6**(4): p. 045012.
17. Rayleigh, L., *XII. On the resultant of a large number of vibrations of the same pitch and of arbitrary phase.* The London, Edinburgh, and Dublin Philosophical Magazine and Journal of Science, (1880). **10**(60): p. 73-78.
18. Nakagami, M., *The m-distribution—a general formula of intensity distribution of rapid fading*, in *Statistical Methods in Radio Wave Propagation*, W.C. Hoffman, Editor. 1960, Pergamon. p. 3-36.
19. Deffieux, T., G. Montaldo, M. Tanter, and M. Fink, *Shear wave spectroscopy for in vivo quantification of human soft tissues visco-elasticity.* IEEE Trans Med Imaging, (2009). **28**(3): p. 313-322.
20. Bhatt, M., M.A.C. Moussu, B. Chayer, F. Destrempes, M. Gesnik, L. Allard, A. Tang, and G. Cloutier, *Reconstruction of viscosity maps in iltrasound shear wave elastography.* IEEE Trans Ultrason Ferroelectr Freq Control, (2019).
21. Parker, K.J., *Power laws prevail in medical ultrasound.* Phys Med Biol, (2022). **67**(9): p. 09TR02.
22. Baek, J., L. Basavarajappa, R. Margolis, L. Arthur, J. Li, K. Hoyt, and K.J. Parker, *Multiparametric ultrasound imaging for early-stage steatosis: Comparison with magnetic resonance imaging-based proton density fat fraction.* Med Phys, (2024). **51**(2): p. 1313-1325.
23. Baek, J., A.M. O'Connell, and K.J. Parker, *Improving breast cancer diagnosis by incorporating raw ultrasound parameters into machine learning.* Mach Learn Sci Technol, (2022). **3**(4): p. 045013.